\documentclass[aps, reprint]{revtex4-1}
\usepackage{graphicx}
\usepackage{physics}

\begin{document}
\title{Quantum Limits to Classically Spoofing an Electromagnetic Signal}
\author{Jonathan N. Blakely}
\affiliation{U. S. Army Combat Capabilities Development Command Aviation \& Missile Center, Redstone Arsenal, Alabama 35898, USA}
\author{Shawn D. Pethel}
\affiliation{U. S. Army Combat Capabilities Development Command Aviation \& Missile Center, Redstone Arsenal, Alabama 35898, USA}

\begin{abstract}
Spoofing an electromagnetic signal involves measuring its properties and preparing a spoof signal that is a close enough copy to fool a receiver. A classic application of spoofing is in radar where an airborne target attempts to avoid being tracked by a ground-based radar by emitting pulses indicating a false range or velocity. In certain scenarios it has been shown that a sensor can exploit quantum mechanics to detect spoofing at the single-photon level.  Here we analyze an idealized spoofing scenario where a transmitter-receiver pair, seeking to detect spoofing, utilizes a signal chosen randomly from a set of non-orthogonal, coherent states. We show that a spoofer optimally employing classical information on the state of the transmitted signal (i.e. the best measure-and-prepare strategy allowed by quantum mechanics) inevitably emits imperfect spoofs that can be exploited by the receiver to reveal the presence of the spoofer, or to discriminate between true reflections and spoofs. Importantly, we show that the quantum limitations on classical spoofing remain significant even in the large mean-photon-number regime.
\end{abstract}
\maketitle
\section{Introduction}
Many remote sensing technologies operate by observing the reflection of a transmitted electromagnetic pulse off of an object of interest. Information about the object (e.g.  range, velocity, orientation, or identity) is derived from properties of the returned pulse (e.g. time of flight, phase, Doppler shift, or polarization). Other properties (e.g. pulse shape or spectral content) may be used by the receiver to distinguish the reflected signal from interference and background noise. This process is sometimes required to operate under conditions where an adversary attempts to fool the receiver by emitting a spoof of the sensing signal prepared with properties indicating a false  range, velocity, orientation, etc. \cite{schleher1999electronic, genova2018electronic}. At the same time, the properties used by the receiver to pick the pulse out of the background noise must be faithfully reproduced in the pulse prepared by the spoofer. Assuming the spoofer does not have complete knowledge of these properties in advance, they must be determined through measurement of the transmitted signal. A classic application of spoofing is where an airborne target emits spoof pulses to avoid being tracked by a ground-based radar \cite{schleher1999electronic, genova2018electronic}. Spoofing also has non-adversarial applications in hardware-in-the-loop testing \cite{strydom2012hardware, strydom2014high, heagney2018digital}. It is therefore interesting to investigate the physical limits on the effectiveness of spoof signals and a receiver's ability to detect them.

Practically speaking, the performance of existing spoofing technologies following a ``measure and prepare'' strategy is limited by thermal phenomena giving rise to various forms of noise \cite{schleher1999electronic, genova2018electronic}. However, technological evolution tends to provide better and better noise mitigation methods over time. In contrast, quantum mechanics imposes a hard limit on the information that can be gained by the spoofer through a measurement on the transmitted pulse. A spoof pulse that fails to resemble the transmitted signal closely enough can make the spoofer vulnerable to detection. This situation is similar to that of quantum cryptography where an eavesdropper interferes with communication between Alice and Bob, but here the transmitter-receiver pair needs only to detect the eavesdropper (i.e. the spoofer) and does not attempt to generate and share a secret key \cite{pirandola2020advances}. Some proposals for exploiting quantum effects to detect spoofing have already appeared in the literature. Detection of spoofing based on non-orthogonal, single-photon polarization states of optical fields was previously considered in the context of  perimeter security \cite{humble2009sensing},  secure imaging \cite{malik2012quantum}, and tamper detection \cite{williams2016tamper}. A different approach to spoof detection utilized entangled pairs of photons in secure lidar \cite{zhao2021quantum}. 

The goal of this paper is to consider more broadly the limits to spoofing imposed by quantum mechanics, going beyond single photon states. To that end, we analyze an idealized spoofing scenario framed as a quantum hypothesis test performed by a receiver to detect a possible spoofer.  We assume the transmitter chooses a signal at random from a predefined signal set containing pulses with non-orthogonal quantum states. The randomly chosen signal is known at the receiver. The spoofer has complete knowledge of the signal set, but not of which specific signal is transmitted. We also assume the spoofer has prior knowledge of, or the means to determine, a complete classical description of the transmitted signal including, for example, its spatio-temporal shape, frequency content, and polarization. What the spoofer does not have is full knowledge of the quantum state of the pulse, which must be determined from a measurement.  

The spoofer prepares the quantum state of the spoof pulse according to the measurement outcome; thus, we assume a ``classical'' spoofer in the sense of one who employs only classical information (i.e. a single measurement outcome) to prepare the quantum state of the spoof signal. This usage is adapted from the literature on benchmarking quantum key distribution schemes \cite{hammerer2005quantum, chiribella2014quantum}. A ``quantum'' spoofer, by contrast, would exploit more of the information in the quantum state than is obtained from a direct measurement.

We do not specify a particular architecture for the receiver and the spoofer. Rather, the performance of both the spoofer's measurement and the receiver are assumed to be as good as possible without violating the laws of quantum mechanics. Finally, we neglect all influence of loss and noise. The end result is not a design for a sensing or spoofing technology, but rather a quantification of the fundamental limits of performance allowed by quantum mechanics.

Under these assumptions, we show that the inability of a classical spoofer to perfectly discriminate non-orthogonal states means the spoof pulses inevitably contain errors that give the receiver a basis for detecting the influence of the spoofer. We  demonstrate this quantum effect is not limited to low mean-photon-number states, nor is it dependent on one specific framing of the spoofing problem. These results motivate further research into realizable demonstrations of quantum-limited, classical spoofing, as well as consideration of the possibilities open to a fully ``quantum'' spoofer.

\section{Spoofing a Binary Phase-Shift Keying Signal}
\label{BPSKsignal}
Our idealized spoofing scenario is intended to retain only the essential elements of a realistic spoofing application needed to display quantum effects. We begin by assuming the transmitter emits a signal represented by a coherent state of a single, pulse-shaped, generalized temporal mode \cite{ou2017quantum}. The complex amplitude of the coherent state is randomly chosen with equal probability to be either $+\alpha$ or $-\alpha$, i.e. a binary phase-shift keying signal \cite{burenkov2021practical}. Importantly, these two states are not orthogonal as $\bra{\alpha}\ket{-\alpha} = e^{-2 N}$, where  $N = \left | \alpha \right |^2$ is the mean photon number for either state.

The spoofer intercepts this signal.  Presumably some property of the spoof signal such as the timing or Doppler shift is chosen deliberately to mislead the receiver. But the quantum state of the spoof pulse must be as faithful a copy of the original as possible in order to be accepted by the receiver as a true reflected pulse. Thus, the spoofer needs to make the measurement that best discriminates between two coherent states with opposite phase and equal prior probability.  The optimal measurement has been shown to successfully discriminate with probability $\gamma$ where \cite{helstrom1969quantum}
\begin{align}
\gamma = \frac{1}{2}\left( 1 +\sqrt{1-e^{-4N}}\right).
\label{BPSK_spoofer_meas_prob}
\end{align}
The spoofer prepares a new coherent state depending on the discrimination outcome, which is then directed back at the receiver.  

When a pulse arrives at the receiver, a decision must be made between two hypotheses. Without loss of generality, suppose the transmitted signal has amplitude $+\alpha$. The first hypothesis, $H_0$, holds that the received signal is a true reflection from the target, so the quantum state of the received signal is represented by the density operator
\begin{figure}[bht]
\includegraphics{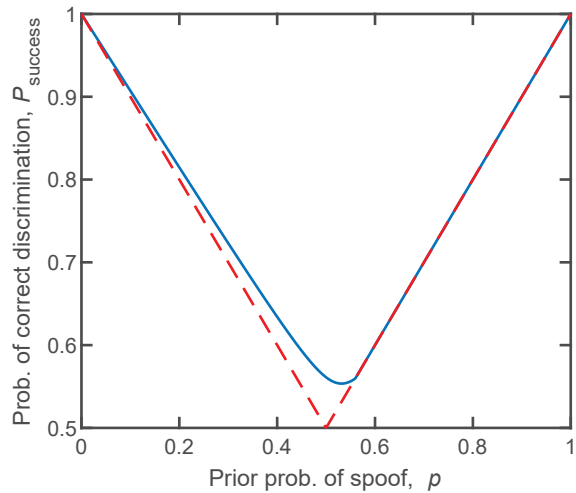}
\caption{Probability of success at discriminating spoofs using binary phase-shift keying with a prior probability of spoofing, $p$, and mean photon number $N = 0.1$. The dotted red line corresponds to a perfect spoofer, i.e. $\gamma = 1$. The solid blue line corresponds to a spoofer making the optimal quantum measurement.}
\label{fig:p_vs_P_success}
\end{figure}\begin{align}
\hat{\rho}_0 = \ket{ \alpha}\bra{ \alpha}.
\label{H_0_density_operator}
\end{align}
The second hypothesis, $H_1$, states that the signal is generated by a spoofer. The receiver assumes the spoofer optimally discriminates the transmitted state, but does not have knowledge of the spoofer's measurement outcome. As a result, the state of the spoofed signal is a statistical mixture represented by the density operator
\begin{align}
\hat{\rho}_1 = \gamma  \ket{ \alpha}\bra{ \alpha} + (1-\gamma)  \ket{ -\alpha}\bra{ -\alpha},
\label{H_1_density_operator}
\end{align}
where $\gamma$ is given by Eq. \ref{BPSK_spoofer_meas_prob}.

If the prior probability of $H_1$ is $p$ and $H_0$ is $1-p$, then the maximum probability of success allowed by quantum mechanics in discriminating a true reflection from a spoof is \cite{1976quantum}
\begin{align}
P_\text{success}= \frac{1}{2}\left ( 1+ ||p \hat{\rho}_1 - (1-p) \hat{\rho}_0 ||_1\right ),
\label{Helstrom}
\end{align}
where $||\cdot||_1$ denotes the trace norm. The trace norm of a Hermitian operator equals the sum of the absolute values of its eigenvalues. In Sec. \ref{BPSKeigenvalues} of the Appendix it is shown that the non-zero eigenvalues of $p \hat{\rho}_1 - (1-p) \hat{\rho}_0$ are $\eta_+$ and $\eta_-$, where
\begin{align}
\eta_\pm &= \frac{1}{2}-p \hspace{0.1cm}\pm \label{Eta_pm} \\
& \sqrt{\left(\frac{1}{2}-p \right)^2 - ( p\gamma - 1+p) p(1-\gamma)\left(1-e^{-4N}\right)}. \nonumber
\end{align}
Then the probability of success in detecting the spoof is
\begin{align}
P_\text{success} = \frac{1}{2}\left ( 1+  |\eta_+ | + |\eta_- | \right )
\label{BPSK_success}
\end{align}

\begin{figure}[tbh]
\includegraphics{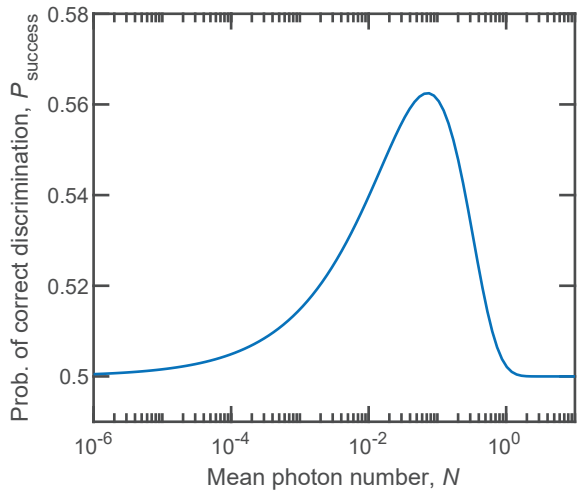}
\caption{Probability of success at discriminating spoofs using binary phase-shift keying as a function of the mean photon number, $N$. The prior probability of a spoof is fixed at $p=1/2$.}
\label{fig:P_success_vs_N}
\end{figure}
To appreciate the significance of Eq. \ref{BPSK_success}, consider the un-physical scenario where the spoofer is able to perfectly discriminate the transmitted quantum state, i.e. $\gamma = 1$. The spoof pulses would always be prepared with exactly the right quantum state. Then the receiver would gain no indication of the spoofer's presence by making measurements on received pulses. The only basis for choosing one hypothesis over the other would be the prior probabilities; thus, the receiver's optimal strategy would be to choose the hypothesis with the largest prior probability. Figure \ref{fig:p_vs_P_success} compares the probability of success under that strategy (dashed red line) with that of Eq. \ref{BPSK_success} (solid blue line) as  functions of $p$, the prior probability of a spoof, in the case of mean photon number $N=0.1$. The separation between these lines manifests the greater susceptibility to detection of a classical spoofer subject to the laws of quantum mechanics compared one who is not.  Interestingly, when $p > 1/(\gamma + 1)$, the optimal strategy is to forego the measurement and assume the pulse is a spoof. Apparently, the quantum limitation vanishes in that regime.

It is also interesting to fix the prior probability of spoofing, $p$, and vary the mean photon number of the signal, $N$. Figure \ref{fig:P_success_vs_N} shows the probability of successful discrimination as a function of the mean photon number with $p=1/2$. The advantage gained over the prior probability peaks at 0.1 photons, and all but vanishes below $10^{-5}$ photons and above $2$ photons. The advantage declines at low photon number since the eigenvalues given in Eq. \ref{Eta_pm} with $p=1/2$ tend toward zero as $N$ approaches zero. The decline at large photon number is due to the decreasing overlap $\bra{\alpha}\ket{-\alpha} = e^{-2 N}$ as $N$ increases.
\begin{figure}[htb]
\includegraphics{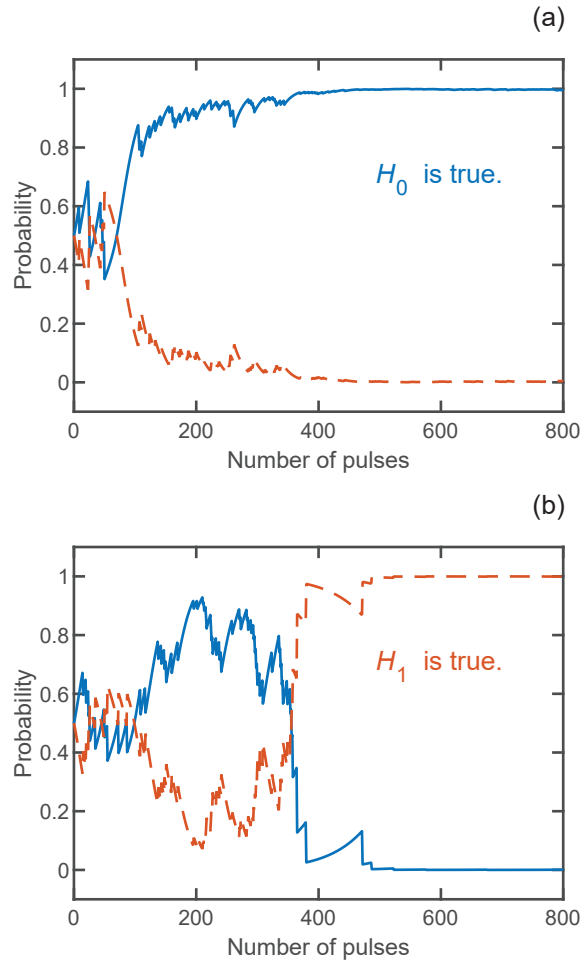}
\caption{Examples of Bayesian inference applied to detecting a spoofer using binary phase-shift keying where (a) no spoofer is present ($H_0$), and (b) a spoofer is present ($H_1$). The solid blue line shows the prior probability of hypothesis $H_0$. The dashed red line shows the prior probability of hypothesis $H_1$.}
\label{fig:Bayes_Updating}
\end{figure}

If the transmitter can send a series of pulses rather than just one, the receiver can incorporate the information gained from successive received signals into updated values of the prior probabilities to better inform the final decision between hypotheses. A common approach is Bayesian inference where an initial value of $p=1/2$ is assumed, and $p$ is updated after each pulse is received \cite{winkler2003introduction}. The updates are derived from expectation values of the density operators $\hat{\rho}_0$ and $\hat{\rho}_1$ in the eigenstate corresponding to the positive eigenvalue $\eta_+$. Details of the calculation are given in Sec. \ref{BayesAppendix} of the Appendix. Two typical instances of this procedure are shown in Fig. \ref{fig:Bayes_Updating} where, first, hypothesis $H_0$ is true (Fig. \ref{fig:Bayes_Updating}(a)), and then $H_1$ is true (Fig. \ref{fig:Bayes_Updating}(b)). The mean photon number is $N=0.1$. In both case, the information gained from measurements drives the process to converge on the true hypothesis. The spoofer's inability to perfectly reproduce the quantum state of the pulse is a vulnerability that enables the receiver to detect it.

In this section, an idealized scenario was devised to contain only the minimum elements of reality needed to expose the difficulty that quantum mechanics poses to a classical spoofer. Next we broaden this result by considering some variations on this scenario that incorporate some small steps toward a realistic application.

\section{Spoofing a Gaussian-Noise Modulated Signal}
\label{GaussianNoiseSignal}
A sensing technology that is essentially restricted to the single photon regime has limited applicability. Figure \ref{fig:P_success_vs_N} shows that for mean photon numbers greater than 2, the binary phase-shift keyed pulse of the previous section offers a vanishingly small advantage over the prior probability of $1/2$. Here we consider an alternative signal set that enables spoof detection at higher mean photon number. The set contains coherent states whose complex amplitude, $\alpha$, is a random value drawn from a Gaussian probability density 
\begin{align}
P(\alpha) = \frac{\lambda}{\pi} e^{-\lambda|\alpha|^2},
\label{CSdensity}
\end{align}
where the width of the density is determined by the constant $\lambda \ge 0$. The case of a flat distribution corresponds to the limit where $\lambda \to 0$. The choice of this class of signals is based primarily on analytical convenience. But a Gaussian-noise modulated waveform for remote sensing is not wholly implausible \cite{narayanan2000doppler, luong2019quantum, luong2020quantum}.

The spoofer intercepts this signal and makes a measurement to gain information about the quantum state in order to reconstruct it as closely as possible.  The ``closeness'' of two quantum states with density operators $\hat{\rho}$ and $\hat{\sigma}$ can be quantified in terms of the fidelity $F(\hat{\rho},\hat{\sigma})$, where $0 \le F(\hat{\rho},\hat{\sigma}) \le 1$, and $ F(\hat{\rho},\hat{\sigma})=1$ if and only if $\hat{\rho}=\hat{\sigma}$ \cite{nielsen2010quantum}. Hammerer\em et al.\em \cite{hammerer2005quantum} analyzed the process of measuring a coherent state  randomly selected from the density in Eq. \ref{CSdensity}, and then using the measurement outcome to reconstruct that state and found the average fidelity $\bar{F}$ of the reconstructed state to the original is bounded by
\begin{align}
\bar{F} \le \frac{1+\lambda}{2+\lambda}.
\end{align}
If $\lambda \ll 1$, then, on average, a spoof prepared by this procedure is a low fidelity reconstruction of the transmitted signal. Importantly, the bound is saturated by heterodyne detection \cite{braunstein2000criteria}; thus, to make the spoof as convincing as possible, we assume here that the adversary makes a heterodyne measurement of the transmitted coherent state.

The outcome $\alpha'$ of an ideal heterodyne measurement on a coherent state of complex amplitude $\alpha$ is a random variable with probability density
\begin{align}
P(\alpha') = \frac{1}{\pi} e^{-|\alpha'-\alpha|^2}.
\label{HMdensity}
\end{align}
Having made the optimal measurement for characterizing the state of the transmitted pulse, the spoofer prepares a coherent state $\ket{\alpha'}$ as a spoof. 

At the receiver, a binary hypothesis test discriminates a true reflection from a spoof. As in Sec. \ref{BPSKsignal}, the first hypothesis, $H_0$, holds that the received signal is a true reflection from the target. In this case, the quantum state of the received signal is represented by the density operator
\begin{align}
\hat{\rho}_0 = \ket{\alpha}\bra{\alpha}.
\label{H_0_density_operator}
\end{align}
The second hypothesis, $H_1$, states that the signal is generated by a spoofer. The receiver  does not know the outcome of the spoofer's measurement. Consequently, the state of the spoofed signal is a statistical mixture represented by the density operator
\begin{align}
\hat{\rho}_1 = \frac{1}{\pi}\int  e^{-|\beta-\alpha|^2} \ket{\beta}\bra{\beta} d^2\beta.
\label{H_1_density_operator}
\end{align}
The maximum probability of success allowed by quantum mechanics in discriminating a true reflection from a spoof is again given by Eq. \ref{Helstrom}.

The trace norm appearing in Eq. (\ref{Helstrom}) is not easy to evaluate given the density operators in Eqs. (\ref{H_0_density_operator}) and (\ref{H_1_density_operator}). However, a lower bound can easily be derived when the prior probability of a spoof $p=1/2$. In that case, the trace norm is subject to the bound \cite{nielsen2010quantum}
\begin{align}
||\frac{1}{2} \hat{\rho}_1 - \frac{1}{2} \hat{\rho}_0 ||_1 \ge F\left( \hat{\rho}_0, \hat{\rho}_1\right)^2,
\label{F_bound}
\end{align}
where the fidelity $F\left( \hat{\rho}_0, \hat{\rho}_1\right) = \sqrt{\bra{\alpha} \hat{\rho}_1\ket{\alpha}} = 2^{-1/2}$. Then $P_\text{success} \ge  3/4$. This probability exceeds the prior probability independent of the value of $\alpha$ or $N$. As a result, the spoofer is vulnerable to detection at any value of the mean photon number.

Further confirmation of this conclusion can be found by numerically approximating the trace norm in the case where $p=1/2$.
To that end, we rewrite the density operators in the number state representation. The number state representation of the coherent state of Eq. (\ref{H_0_density_operator}) is well known  \cite{loudon2000quantum} to have matrix elements
\begin{align}
\bra{m}\hat{\rho}_0\ket{n} = e^{-|\alpha|^2}\frac{\alpha^m}{\sqrt{m!}}\frac{(\alpha^*)^n}{\sqrt{n!}}.
\end{align}
The mixed state of hypothesis $H_1$ takes the form of a noisy coherent state or a displaced thermal state with mean noise photon number of 1. The density operator for such a state has matrix elements \cite{marian1993squeezed} 
\begin{align}
\bra{m}\rho_1 \ket{n}&=\text{exp} \left( -\frac{2|\alpha|^2}{3} \right) \left( \frac{m!}{n!}\right)^{1/2} \alpha^{m-n}  \nonumber \\
& \times \frac{\left(\frac{1}{2}\right)^m}{\left(\frac{3}{2}\right)^{m+1}} L_m^{(m-n)} \left [ -\frac{4|\alpha|^2}{3} \right ],
\end{align}
\begin{figure}[tbh]
\includegraphics{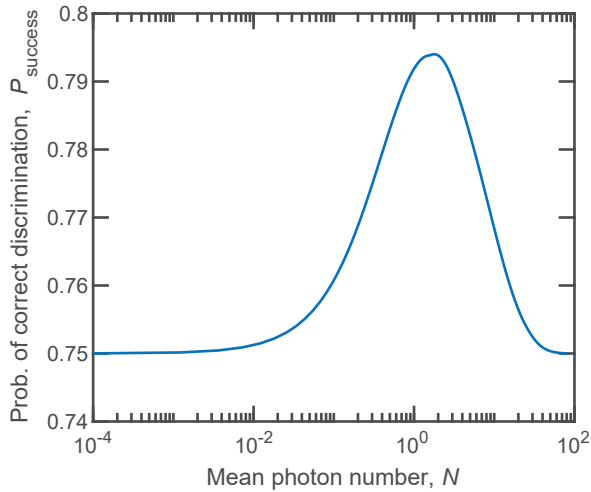}
\caption{Probability of success at discriminating spoofs using Gaussian-noise modulated pulses as a function of the mean photon number, $N$. The prior probability of a spoof is fixed at $p=1/2$.}
\label{fig:P_success_vs_N_Gauss}
\end{figure}where $ L_m^{(m-n)} $ is a Laguerre polynomial. Using these matrix elements, the trace norm in Eq. \ref{Helstrom} can be numerically approximated by finding the eigenvalues of the operator $p \hat{\rho}_1 - (1-p) \hat{\rho}_0 $ truncated at a maximum photon number far above the mean photon numbers of the states in Eqs.  (\ref{H_0_density_operator}) and (\ref{H_1_density_operator}), i.e.  $|\alpha|^2$ and $|\alpha'|^2$.

Figure \ref{fig:P_success_vs_N_Gauss} shows the probability of successful discrimination of spoofs approximated in this manner as a function of the mean photon number, $N$, in the case where $p=1/2$. The density matrices were truncated at up to 145 photons. Two features of this plot are notable. First, the probability peaks around a mean photon number of 2, an order of magnitude higher than the peak at 0.1 photons in Fig. \ref{fig:P_success_vs_N}. Second, consistent with Eq. \ref{F_bound}, the probability is greater than 0.75 for all values of $N$, and never falls to the prior probability, p = 1/2, at any mean photon number.  

The Gaussian-noise modulated signal set enables detection of the spoofer at any signal power level, and is not limited to the single photon regime because the extent of overlap $\bra{\alpha}\ket{\alpha'} = e^{-\frac{1}{2}|\alpha - \alpha'|^2}$ depends only on the difference $|\alpha - \alpha'|$ and not on the magnitude of $\alpha$ and $\alpha'$ separately. More broadly, this result suggests any signal set that contains close coherent states, regardless of mean photon number, could be used to detect a spoofer.

\section{Discriminating Two Received Pulses}
\label{TwoReturns}
One way in which the spoofing scenarios examined in the preceding sections differ from realistic radar spoofing is that they allow for only a single pulse arriving at the receiver. In practice, the transmitted pulse always reflects off the spoofer to some \begin{figure}[tbh]
\includegraphics{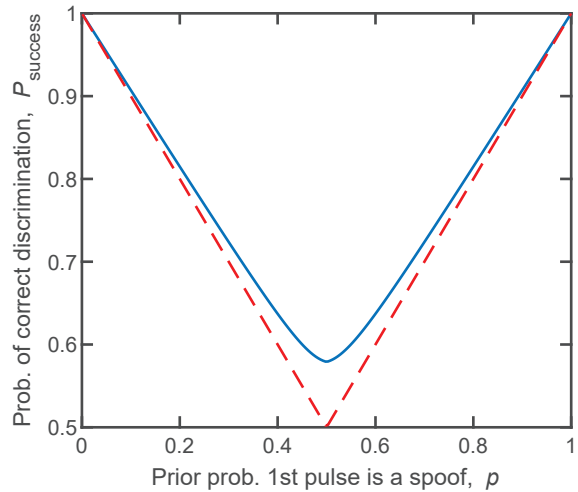}
\caption{Probability of success at discriminating spoofs when two pulses are received as a function of the prior probability of the first pulse being a spoof, $p$. The dotted red line corresponds to the case of a perfect spoofer, i.e. $\gamma = 1$. The solid blue line corresponds to a spoofer making the optimal quantum measurement. In both cases, the mean photon number in the signal pulse is $N = 0.1$}
\label{fig:p_vs_P_skin}
\end{figure}extent. This pulse is referred to as the ``skin return'' \cite{schleher1999electronic, genova2018electronic}. The spoofer emits a separate spoof pulse. So two pulses actually arrive at the receiver. The receiver must decide which is which. This scenario requires a different hypothesis test at the receiver, and we will show that quantum mechanics again exposes the spoofer to detection.

We return to the binary phase-shift keying signal set of Sec. \ref{BPSKsignal}. Assuming two pulses are received, under either hypothesis the density operator contains two modes, one for each pulse. Hypothesis $H_0$ supposes the first mode is the skin return, with single-mode density operator
\begin{align}
\hat{\rho}_\text{ret} =  \ket{ \alpha}\bra{ \alpha},
\end{align}
 and the second is the spoof, with single-mode density operator
 \begin{align}
 \hat{\rho}_\text{sp} = \gamma  \ket{ \alpha}\bra{ \alpha} + (1-\gamma)  \ket{ -\alpha}\bra{ -\alpha}.
 \end{align}
Then the density operator for the two modes received under $H_0$ is
\begin{align}
\hat{\rho}_0=\hat{\rho}_\text{ret} \otimes\hat{\rho}_\text{sp}.
\label{Skin_0}
\end{align} 
 Hypothesis $H_1$ supposes the converse. So the density operator for the two modes received under $H_1$ is
\begin{align}
\hat{\rho}_1  = \hat{\rho}_\text{sp}\otimes\hat{\rho}_\text{ret}.
\label{Skin_1}
\end{align}
Again, the eigenvalues of  $p \hat{\rho}_1 - (1-p) \hat{\rho}_0$ can be solved for, and the probability of successful discrimination determined through Eq. \ref{Helstrom} (see Sec. \ref{TwoRetEigen} of the Appendix for details). This probability is plotted in Fig. \ref{fig:p_vs_P_skin} (blue line) along with maximum prior probability (dashed red line) as functions of $p$, as was done in Fig. \ref{fig:p_vs_P_success}. Interestingly, the success probability in Fig. \ref{fig:p_vs_P_skin}  is symmetric about $p=1/2$, in contrast to the probability in Fig. \ref{fig:p_vs_P_success}. This symmetry can be seen explicitly in the density operators of Eqs. \ref{Skin_0} and \ref{Skin_1}, but not in Eqs. \ref{H_0_density_operator} and \ref{H_1_density_operator}. More importantly, the separation between the curves indicates the information gained from the measurement that would enable detection of the spoofer, for example using Bayesian inference as in Sec. \ref{BPSKsignal}. Thus, the vulnerability to detection of classical spoofing identified in Secs. \ref{BPSKsignal} and \ref{GaussianNoiseSignal} is not limited to one specific framing of the problem.

\section{Discussion}
By analyzing idealized spoofing scenarios, we have shown that a transmitter-receiver pair choosing randomly from a set of coherent states can detect the presence of a spoofer because quantum mechanics precludes the optimum spoofer from identifying the transmitted signal with certainty. Importantly, when transmitted signals are drawn from a set of coherent states with Gaussian distributed amplitudes, successful detection of spoofing is not limited to a low power regime. 
 
Several extensions to this work are possible. Effects of noise and loss can be introduced to obtain a more practical assessment of the quantum limits on classical spoofing. These effects are especially prominent in the microwave regime where reflections are weak and thermal noise is strong. The optimal receivers considered here can be replaced by specific receiver architectures. Alternative signal sets can be analyzed to facilitate generation, detection, or other aspects of system performance. Our results should motivate experiments designed to probe the limits of classical spoofing in various regimes of the electromagnetic spectrum. An optical implementation would probably be the most straightforward for first proof-of-principle, perhaps along the lines of recent free-space, continuous-variable quantum key distribution demonstrations \cite{Qu:17, shen2019free}.

It is likely that the limits on classical spoofing examined here can be overcome by considering ``quantum'' spoofing. For example, perhaps a pulse can be coupled into a cavity, modified coherently, and re-emitted with the quantum state intact but with a shifted frequency or polarization. A more exotic approach would be to teleport the quantum state of the incoming pulse onto a spoof pulse \cite{fedorov2021experimental}. Future generations of remote sensing technology design may require consideration of such possibilities.

\section{Acknowledgment}
The authors would like to thank Kurt Jacobs and Martin S. Heimbeck for helpful discussions of this work.

\appendix*
\section{Additional Details}
\subsection{Eigenvalues in Sec. \ref{BPSKsignal}}
\label{BPSKeigenvalues}
The probability of successful discrimination in Eq. \ref{Helstrom} can be expressed in terms of the non-zero eigenvalues of the operator
\begin{align}
p \hat{\rho}_1& - (1-p) \hat{\rho}_0  =\\
 p&  \Big [ \gamma  \ket{ \alpha}\bra{ \alpha} + (1-\gamma)  \ket{ -\alpha}\bra{ -\alpha} \Big ]- (1-p) \ket{ \alpha}\bra{ \alpha}\nonumber 
\end{align}
An eigenvector $\ket{\eta}$ corresponding to the non-zero eigenvalue $\eta$ must lie in the subspace spanned by $\ket{ \alpha}$ and $\ket{ -\alpha}$ and therefore may be written as
\begin{align}
\ket{\eta} = c_+ \ket{\alpha} + c_- \ket{-\alpha}
\label{eigenvec_form}
\end{align}
where $c_+$ and $c_-$ are complex coefficients. The corresponding eigenvalue $\eta$ satisfies
\begin{align}
\left |\begin{matrix}
& p\gamma - (1-p) -\eta  &  (p\gamma - (1-p))e^{-2N} \\
&p(1-\gamma)e^{-2N}  & p(1-\gamma) - \eta 
\end{matrix}
\right |=0.
\end{align}
The two solutions are given in Eq. \ref{Eta_pm}.

\subsection{Bayesian Inference in Sec. \ref{BPSKsignal}}
\label{BayesAppendix}
Optimal quantum discrimination between two hypotheses involves a measurement operator that projects onto the positive subspace of the operator $p \hat{\rho}_1 - (1-p) \hat{\rho}_0$ and has eigenvalues 0 and 1 corresponding to a measurement outcome indicating a decision in favor of the corresponding hypothesis. In the current instance, $\eta_+$ as defined by Eq. 5 is the unique positive eigenvalue. So the optimal measurement is a projection onto the corresponding eigenvector $\ket{\eta_+}$. Then, given this eigenvector, the probability $P(i|H_j)$  of getting a measurement outcome $i$ given that hypothesis $H_j$ is correct, where $i, j \in \{0, 1\}$, can be determined. These probabilities enable the process of Bayesian inference whereby the prior probabilities are updated with each new measurement to reflect new information gained about the truth or falsehood of the respective hypotheses. We first explain the rules for updating the prior probabilities given $P(i|H_j)$, and then explain how to detemine $P(i|H_j)$ itself.

Bayesian inference involves updating the prior probability after each new measurement outcome as follows \cite{winkler2003introduction}. Let $P(H_i)$ be the prior probability of hypothesis $H_i$ where $i \in \{0, 1\}$. For example, in the previous sections, $P(H_1)=p$. After a measurement is made, the posterior probability $P(H_i|j)$  for $H_i$ given the measurement outcome of $j$ incorporates the new information gained about the truth of the hypotheses. If a measurement results in outcome $1$, 
\begin{align}
P(H_0|1) &=\frac{P(1|H_0)P(H_0)}{P(1|H_1)P(H_1) + P(1|H_0)P(H_0)}, \\
P(H_1|1) &=1-P(H_0|1). 
\end{align}
If a measurement results in an outcome of $0$,
\begin{align}
P(H_0|0) &=\frac{P(0|H_0)P(H_0)}{P(0|H_1)P(H_1) + P(0|H_0)P(H_0)}, \\
P(H_1|0) &=1-P(H_0|0). 
\end{align}
The posterior probabilities are then adopted as the new prior probabilities until yet another measurement is made.

The measurement probabilities $P(i|H_j)$ are determined by the eigenvector corresponding to $\eta_+$, the unique positive eigenvalue specified by Eq. \ref{Eta_pm}. Assuming $\ket{\eta_+}$ takes the form of Eq. \ref{eigenvec_form}, the coefficients $c_+$ and $c_-$ satisfy
\begin{align}
\left (\begin{matrix}
& p\gamma - (1-p) -\eta_+  &  (p\gamma - (1-p)) e^{-2N} \\
&p(1-\gamma) e^{-2N}  & p(1-\gamma) - \eta_+ 
\end{matrix}
\right )\left(\begin{matrix} & c_+ \\ &c_- \end{matrix} \right)=0
\end{align}
Then, after normalization, 
\begin{align}
\ket{\eta_+} =\left(\begin{matrix} & c_+ \\ &c_- \end{matrix} \right)= \left( \begin{matrix} & \frac{1}{\sqrt{1+\left|\frac{\eta_+ - p\gamma + (1-p)}{ (p\gamma - (1-p)) e^{-2N}}\right|^2}} \\
&\frac{1}{\sqrt{1+\left|\frac{ (p\gamma - (1-p)) e^{-2N}}{\eta - p\gamma + (1-p)}\right|^2}} \end{matrix} \right)
\end{align}

The measurement outcome probabilities are
\begin{align}
P&(1|H_0) = \bra{\eta_+}\hat{\rho}_0\ket{\eta_+}\\
&= | c_{+}|^2+ c_{-}^* c_{+} e^{-2N}+  c_{+}^* c_{-} e^{-2N}+ | c_{-}|^2 e^{-4N} \\
P&(1|H_1) = \bra{\eta_+}\hat{\rho}_1\ket{\eta_+} \\
&= \gamma P(1|H_0) + (1-\gamma) \left(c_{+}^*  e^{-2N}+ c_{-}^*\right) \left(c_{+}  e^{-2N}+ c_{-}\right) \\
P&(0|H_0) = 1-P(1|H_0) \\
P&(0|H_1) = 1-P(1|H_1) 
\end{align}

\subsection{Eigenvalues in Sec. \ref{TwoReturns}}
\label{TwoRetEigen}
The probability of successful discrimination in Eq. \ref{Helstrom} can be expressed in terms of the non-zero eigenvalues of the operator $p \hat{\rho}_1 - (1-p) \hat{\rho}_0$, where $\hat{\rho}_0$ is given in Eq. \ref{Skin_0}, and $\hat{\rho}_1$ is given in Eq. \ref{Skin_1}. Since these operators contain only two states, $\ket{\alpha}$ and $\ket{-\alpha}$, and we are only interested in the non-zero eigenvalues, we can assume the corresponding eigenvectors lie in the subspace spanned by tensor products of the states $\ket{\alpha}$ and $\ket{-\alpha}$. Any eigenvectors outside this subspace could only correspond to zero eigenvalues since they would be orthogonal to  $\ket{\alpha}$ and $\ket{-\alpha}$. In fact, we don't need $\ket{-\alpha}\ket{-\alpha}$ because it doesn't appear in the density operator anywhere; thus, we will assume the eigenvectors are of the form
\begin{align}
\ket{\eta} = c_{++} \ket{\alpha}\ket{\alpha} +  c_{+-} \ket{\alpha}\ket{-\alpha} +c_{-+} \ket{-\alpha}\ket{\alpha} 
\end{align}
Non-zero eigenvalues satisfy the equation
\begin{align}
\left |\begin{matrix} 
(2p-1) \gamma   -\eta  &  (2p-1) \gamma e^{-2N}     &  (2p-1) \gamma e^{-2N}  \\
(p-1)(1-\gamma)e^{-2N} & (p-1)(1-\gamma)  -\eta   &(p-1)(1-\gamma)  e^{-4N}   \\
p(1-\gamma)e^{-2N}    &p(1-\gamma) e^{-4N}  & p(1-\gamma)-\eta 
\end{matrix} \right |\nonumber\\
=0
\label{QIcov_0}
\end{align}
This determinant can be expressed as the cubic equation
\begin{align}
a_3 \eta^3 + a_2 \eta^2 + a_1 \eta + a_0 = 0
\label{Cubic}
\end{align}
where
\begin{align}
a_3 &= -1, \\
a_2 &=2p-1, \\
a_1 &=  (2p-1)^2 \gamma (1-\gamma) \left( e^{-4N}-1\right) \nonumber \\
&- p(1-p)(1-\gamma)^2\left(  e^{-8N} -1\right),  \\
a_0 &= -   (2p-1) \gamma p(1-p)(1-\gamma)^2 \left( e^{-4N}-1\right )^2.
\end{align}
Equation \ref{Cubic} can be solved analytically or numerically to obtain non-zero values of $\eta$.

\bibliography{Quant_Limits_Class_Spoof}
\end{document}